\documentclass[a4paper,12pt]{article}
\usepackage[utf8x]{inputenc}
\usepackage{epsfig,amssymb,amsmath,latexsym,color,pifont,graphicx,psfrag,abstract}
\newcommand{\h}{h_{j}}
\newcommand{\f}{f_{i}}
\newcommand{\R}{\mathbf{R}}
\newcommand{\A}{\mathbf{A}}

\newcommand{\E}{\mathbf{E}}
\newcommand{\la}{\langle}
\newcommand{\ra}{\rangle}
\title{\textbf{Network topology reconstructed from derivative-variable correlations}}
\author{Zoran Levnaji\'c\footnote{Correspondence and requests for materials should be addressed to Zoran Levnaji\'c (zoran.levnajic@fis.unm.si)}, Arkady Pikovsky \\ 
Department of Physics and Astronomy, University of Potsdam \\ 14476 Potsdam, Germany }
\date{}

\begin{document}
\maketitle

\begin{abstract}
A method of network reconstruction from the dynamical time series is introduced, relying on the concept of derivative-variable correlation. Using a tunable observable as a parameter, the reconstruction of any network with known interaction functions is formulated via simple matrix equation. We suggest a procedure aimed at optimizing the reconstruction from the time series of length comparable to the characteristic dynamical time scale. Our method also provides a reliable precision estimate. We illustrate the method's implementation via elementary dynamical models, and demonstrate its robustness to both model and observation errors. \\ \\
\end{abstract}

\section*{Introduction}
The development of methods for reconstructing the topologies of real networks
from the observable data, is of great interest in modern network science.
Topology, in combination with the inter-node interactions, determine the
function of complex networks~\cite{barrat}. Reconstruction methods are often
developed within the contexts of particular fields, relying on domain-specific
approaches. These include gene regulations~\cite{hecker,frank,nelson,pigolotti}, metabolic
networks~\cite{herrgard}, neuroscience~\cite{bullmore}, or social
networks~\cite{eagle}. On the other hand, theoretical reconstruction concepts
are based on paradigmatic dynamical models such as phase
oscillators~\cite{us,kralemann,luce,ren}, some of which have been experimentally
tested~\cite{blaha,stankovski}. In a similar context, techniques for detecting
hidden nodes in networks are being investigated~\cite{su}. A class of general
reconstruction methods exploit the time series obtained by quantifying the
network behaviour. Some of them assume the knowledge of the internal interaction
functions~\cite{shandilya,dzeroski}, while others do not~\cite{hempel}. Network
couplings can be examined via information-theoretic approach~\cite{pompe}.
Advantage of these methods is that they are \textit{non-invasive}, i.e. require
no interfering with the on-going network dynamics.

Reconstruction methods are often based on examining the inter-node
correlations~\cite{ren}. On the other hand, universal network models such as
Eq.\ref{eq-1}, are based on expressing the time derivative of a node as a
combination of a local and a coupling term. Inspired by this, we propose a
non-invasive reconstruction method, departing from the concept of
\textit{derivative-variable correlation}. Our method assumes the dynamical time
series to be available as measured observables, and the interaction functions to
be known. We present our theory in a general form, extending our initial
results~\cite{preliminary}. As we show, our approach allows for the
reconstruction precision to be estimated, indicating the level of noise in the
data, or possible mismatches in the knowledge of the interaction functions.

\section*{Results}

\subsection*{The reconstruction method}
We consider a network of $N$ nodes, described by their dynamical states $x_i
(t)$. Its time evolution is governed by: 
\begin{equation} \dot x_i = \f (x_i) + \sum_{j=1}^N   A_{ji}  h_{j}  (x_j)  \, ,
\label{eq-1} \end{equation}
where the function $\f$ represents the local dynamics for each node, and $h_{j}$
models the action of the node $j$ on other nodes. The network topology is
encoded in the adjacency matrix $A_{ji}$, specifying the strength with which the
node $j$ acts on the node $i$. We assume that: (\textit{i}) the interaction
functions $\f$ and $h_j$ are precisely known, and (\textit{ii}) a discrete
trajectory consisting of $L$ values $x_i(t_1),\hdots, x_i(t_L)$ is known for
each node. The measurements of $x_i$ are separated by the uniform observation
interval $\delta_t$ defining the time series resolution. We seek to reconstruct
the unknown adjacency matrix $A_{ij} \equiv \A$ under these two assumptions.

The starting point is to define the following correlation matrices, using the
observable $g(x)$ whose role will be explained later:
\begin{equation} \begin{array}{lll}
\mathbf{B} &=& \langle g(x_i) \dot x_j \rangle  \; ,  \\ 
\mathbf{C} &=& \langle g(x_i) f_j (x_j) \rangle  \; , \\  
\mathbf{E} &=& \langle g(x_i) h_j (x_j) \rangle  \; ,       \label{eq-2}
\end{array} \end{equation}
where $\la \cdot \ra$ denotes time-averaging $\la r \ra = \frac{1}{L}
\sum_{m=1}^L r (t_m)$. Inserting into the Eq.\ref{eq-1}, we obtain the
following linear relation between the correlation matrices:
\begin{equation}  \mathbf{A} = \mathbf{E}^{-1}  \cdot( \mathbf{B} - \mathbf{C} )
\; , \label{eq-3} \end{equation}
which is our main reconstruction equation, applicable to any network with
dynamics given by Eq.\ref{eq-1}. Time series are to be understood as the
available observables, allowing for matrices in Eq.\ref{eq-2} to be computed
for any $g$. For the infinitely long dynamical data, reconstruction is always
correct for any generic $g$. For short time series, representing experimentally
realistic scenarios, the reconstruction is always approximate, and its precision
crucially depends on the choice of $g$~\footnote{Usually, correlations are
defined as central moments with averages subtracted. Instead, we are here not
interested in correlations per se, but in the reconstruction according to
Eq.\ref{eq-3}, for which the subtraction of averages is not needed.}.

To be able to quantify the reconstruction precision, we need to equip ourselves
with the adequate measures. To differentiate from the original adjacency matrix
$\A$, we term the reconstructed matrix $R_{ij} \equiv \R$, and express the
matrix error as:
\begin{equation} \Delta_A = \sqrt{  \frac{   \sum_{ij} ( R_{ij} - A_{ij} )^2}{
\sum_{ij}  A_{ij}^2 } } \, . \end{equation}
Of course, each $\R$ is computed according to Eq.\ref{eq-3} in correspondence with the chosen $g$. 
However, since the matrix $\A$ is unknown, we have to introduce another precision
measure, based only on the available data. A natural test for each $\R$ is to
quantify how well does it reproduce the original data $x_i (t_m)$. We apply the
following procedure: start the dynamics from $x_i (t_1)$ and run it using $\R$
until $t=t_2$; denote thus obtained values $y_i (t_2)$; re-start the run from
$x_i (t_2)$ and run until $t=t_3$, accordingly obtaining $y_i (t_3)$, and so on.
The discrepancy between the reconstructed time series $y_i (t_m)$ and the
original $x_i (t_m)$ is an explicit measure of the reconstruction precision,
based solely on the available data. We name it trajectory error $\Delta_T$, and
define it as follows:
\begin{equation} \Delta_T = \frac{1}{N} \sum_{i=1}^N  \sqrt{ \frac{ \big\la (x_i
- y_i)^2 \big\ra }{  \big\la ( x_i - \la x_i \ra )^2 \big\ra }  } \; . 
\end{equation}
Different choices of the observable $g$ lead to different $\R$, with different
precisions expressed through errors $\Delta_T$ and $\Delta_A$. As we show below,
these two error measures are related, meaning that small $\Delta_T$ suggests
small $\Delta_A$. The function $g$ hence plays the role of a tunable parameter,
which can be used to optimize the reconstruction. By considering many $\R$-s
obtained through varying $g$, we can single out $\R$-s with the minimal
$\Delta_T$ to obtain the best reconstruction.

\subsection*{Implementation of the method}
\begin{figure}[!hbt] \centering 
\includegraphics[width=0.7\textwidth]{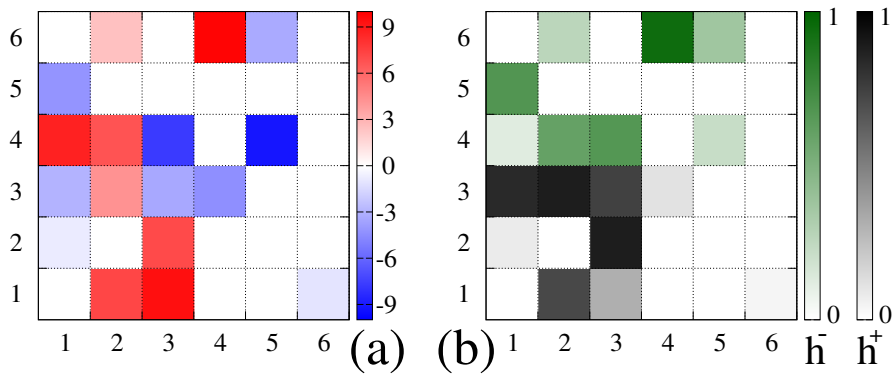}
\caption{Adjacency matrix $\A$ for the first example (a), and the second example
(b). Colorbars (shades) indicate the interaction strength. Two different
colorbars in (b) stand for two different interaction types (see text).}  
\label{figure-1} 
\end{figure}

\begin{figure}[hbt!] \centering 
\includegraphics[width=0.8\textwidth]{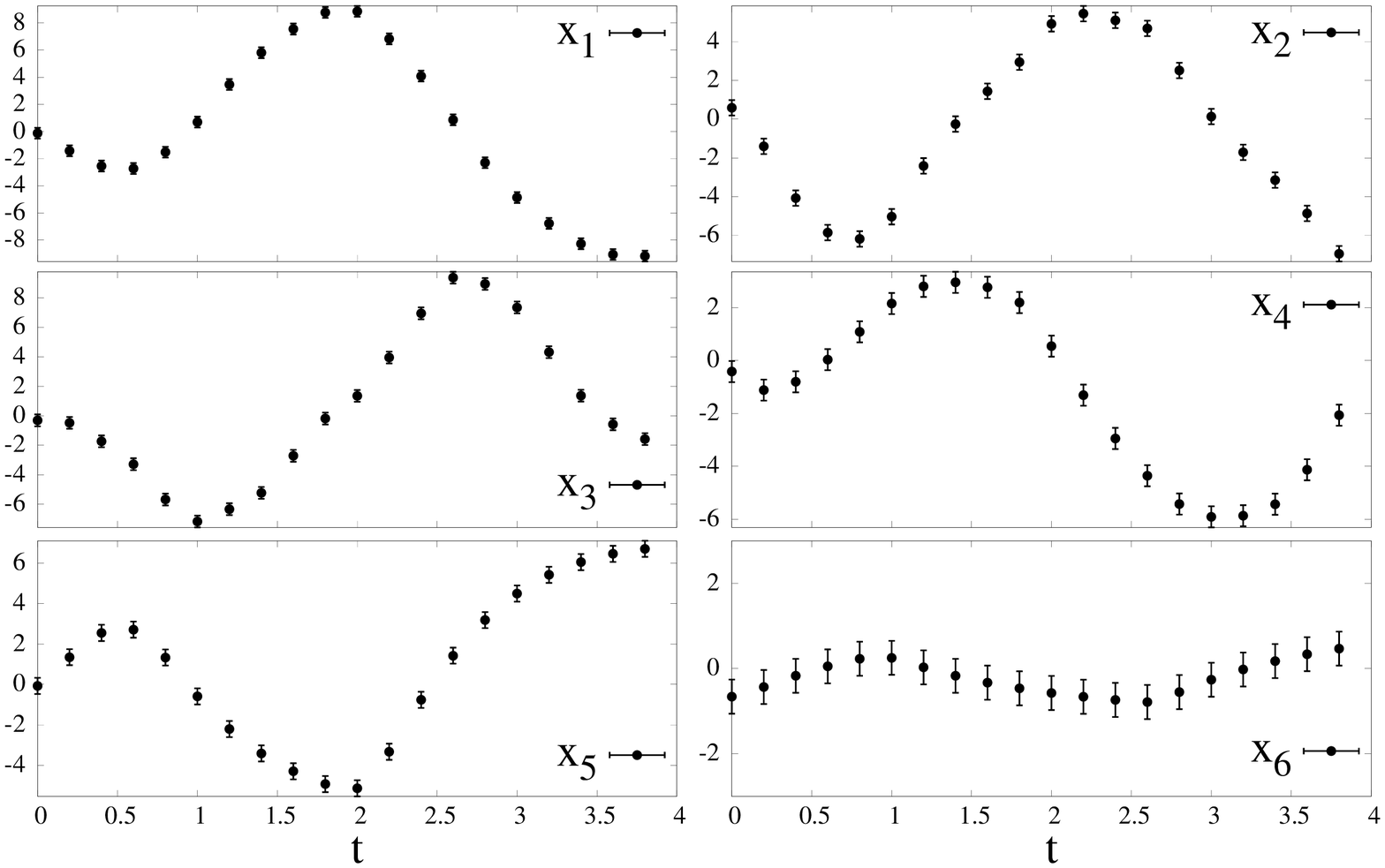} \caption{Time
series for all 6 nodes produced by the network Fig.\ref{figure-1}a (black
dots). Bars denote the added white noise of strength $\eta=0.4$ (see text, cf.
Fig.\ref{figure-5}a).}  \label{figure-2}  
\end{figure}

To illustrate the implementation of our method, we begin by constructing a
network with $N=6$ nodes by putting 17 directed links between randomly chosen
node pairs. As our first example, we consider the Hansel-Sompolinsky model,
describing the firing rates in neural populations~\cite{HS}. It is defined by
the interaction functions $\f = -x$ and $\h = \tanh x$ which are fixed for all
nodes. The adjacency matrix is specified by assigning positive and negative
weights to the networks links, randomly chosen from $[-10,10]$, as shown in
Fig.\ref{figure-1}a.
Starting from random initial conditions, the resulting system is integrated from
$t=0$ to $t=4$. During the run, 20 values of $x_i$ are stored for each node,
equally spaced with $\delta_t=0.2$. The obtained time series, shown in
Fig.\ref{figure-2}, are rather short compared to the characteristic time scale
and the network size. 

\begin{figure}[hbt!] \centering 
\includegraphics[width=0.8\textwidth]{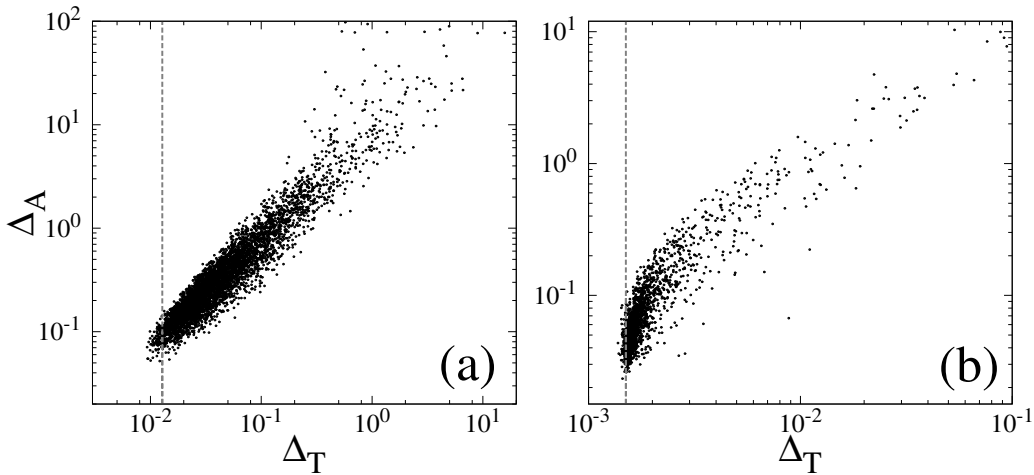} \caption{Scatter
plots of errors $\Delta_T$ and $\Delta_A$, in relation with the first and second
example, in (a) and (b), respectively. Best 1\% of $\R$-s with the minimal
$\Delta_T$, are represented by the dots left of the vertical dashed line.}  
\label{figure-3}   \end{figure}

We now use these data to reconstruct the original adjacency matrix by employing the procedure described above. We consider a set of $10^4$ test-functions $g$,
each composed of first 10 Fourier harmonics
\begin{equation}g(x) = \sum_{k=1}^{10} \big[ a_k \sin(kx) + b_k \cos (kx) \big] \; .  \end{equation}
The coefficients $a_k$ and $b_k$ are randomly selected from $[0,100]$ with the log-uniform probability. This is implemented by selecting each Fourier coefficient via $10^{2.00432137 \times rand} - 1.0$, where $rand$ is a random number between 0 and 1. A typical function thus constructed for each choice of $a_k$ and $b_k$ will have all 10 Fourier components, but one (or at most few) will be well pronounced. Functions are then normalized to the range of time series values. Given relatively smooth timeseries, lower harmonics are expected to generally extract more features from data, which is why we limit ourselves to the first 10 harmonics. To improve the stability of the derivative estimates, we base our
calculations on the set of time points $\tau_m = (t_{m+1} - t_m)/2$. For each
$g$, the matrix $\R$ is obtained via Eq.\ref{eq-2} and Eq.\ref{eq-3}, with the
invertibility of each $\E$ checked by virtue of the singular value
decomposition. The errors $\Delta_T$ and $\Delta_A$ are then calculated for each
$\R$, and reported as a scatter plot in Fig.\ref{figure-3}a.

\begin{figure}[hbt!] \centering 
\includegraphics[width=0.7\textwidth]{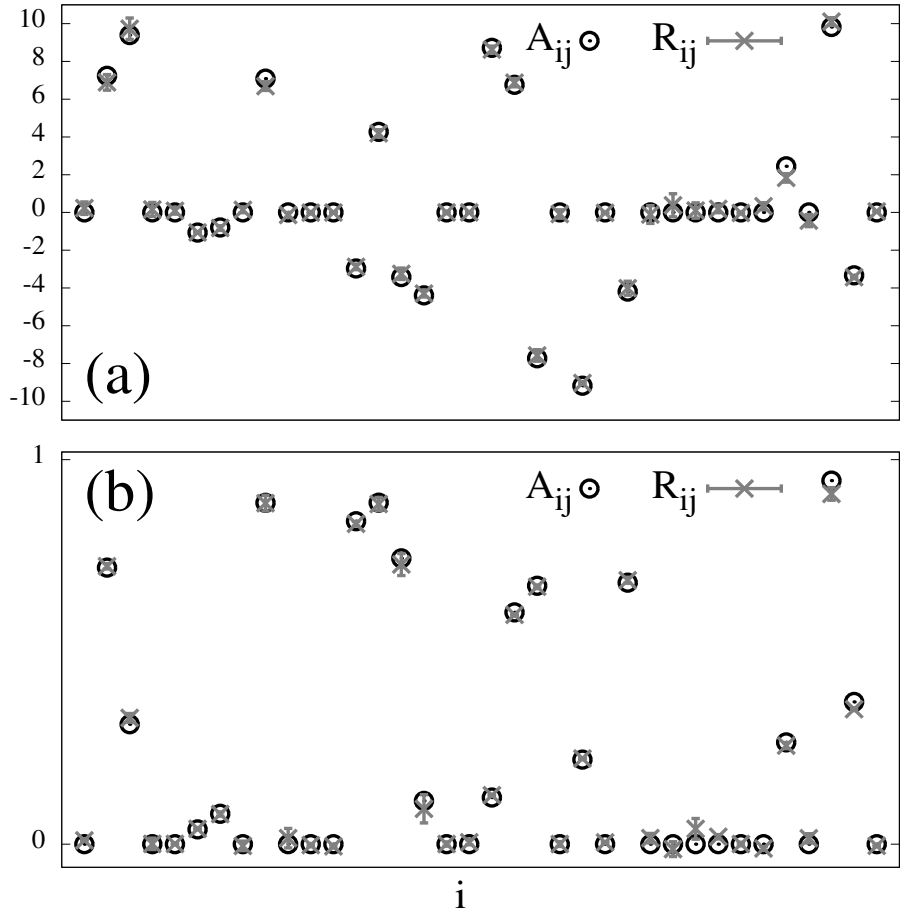}
\caption{Elements of the original $\A$ (circles), and the best reconstruction
(crosses), with the corresponding errorbars. First and second example in (a) and
(b), respectively.}  
\label{figure-4}   \end{figure}

The main result of this analysis is a clear correlation between $\Delta_T$ and
$\Delta_A$, particularly pronounced for smaller values of errors. This confirms
that the best $\R$ are among those that display minimal $\Delta_T$. In order to
identify the best reconstruction and estimate its precision, we focus on the 1\%
of matrices $\R$ with the minimal $\Delta_T$, as illustrated in
Fig.\ref{figure-3}a by the dashed vertical line. The variability of $\R$ within
this group can be viewed as the reconstruction precision. Small variability
indicates the invariance of $\R$ to the choice of $g$, which suggests a good
reconstruction. Large variability of $\R$ implies its drastic dependence on $g$,
indicating a bad precision. We quantify this by computing the mean and the
standard deviation for each matrix element of $\R$ within this group, and
identify them, respectively, with the best reconstruction value and its
precision. In Fig.\ref{figure-4}a we report the original $\A$ and the best
reconstruction, along with the respective errorbar for each matrix element,
describing the reconstruction precision. The reconstruction is indeed very good
for both zero and non-zero weights (i.e. for non-linked and linked node pairs in
the network). 

As our second example, we consider a dynamical model describing gene
interactions, with the coupling functions of two types: activation $h^+_j =
x^5/(1+x^5)$ and repression $h^-_j = 1/(1+x^5)$~\cite{widder}. Local interaction
are again modeled via $f_i=-x$. The adjacency matrix is based on the same
network, and defined by assigning a random weight from $[0,1]$ for each link, as
shown in Fig.\ref{figure-1}b. The nodes 1-3 (respectively, 4-6) act
activatorily (repressively) on all nodes that they act upon. Again, we run the
dynamics from $t=0$ to $t=4$, obtaining another set of time series with 20
points (not shown). The same reconstruction procedure is applied, yielding the
$\Delta_T$ vs. $\Delta_A$ scatter plot shown in Fig.\ref{figure-3}b. Using the
same procedure, we obtain the best reconstruction and show it in
Fig.\ref{figure-4}b. Again, the precision is very good. Note that our
method thus applies also in cases of strongly non-linear interaction functions, which
capture most real dynamical scenarios.

\subsection*{Testing the method's robustness}
\begin{figure}[hbt!] \centering 
\includegraphics[width=0.8\textwidth]{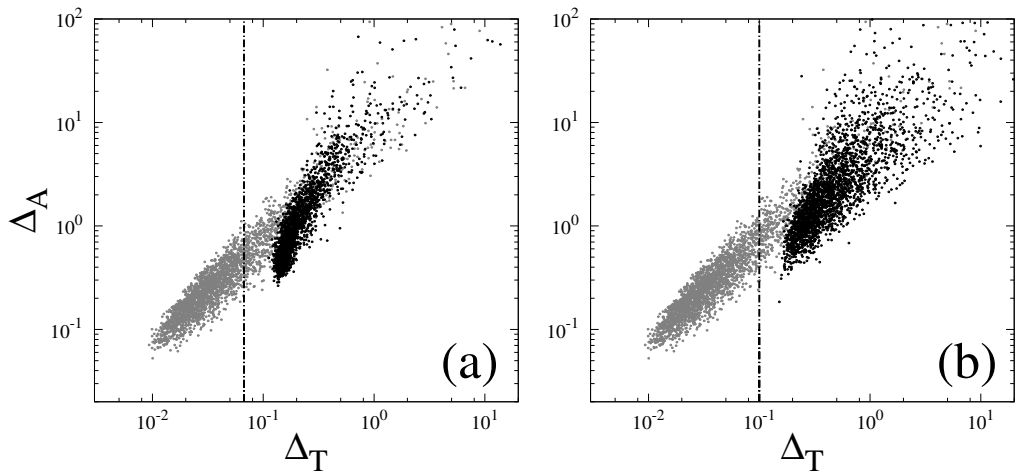} \caption{Scatter
plots of errors $\Delta_T$ and $\Delta_A$ (black dots), for the model error scenario in (a) and the 
observation error scenario in (b). Original non-perturbed scatter plot from Fig.\ref{figure-3}a is shown in 
gray for comparison. Vertical dashed lines depicts the part of the $\Delta_T$ error which is unavoidable in the 
presence of the perturbation (see text).}
\label{figure-5}   \end{figure}

\begin{figure}[hbt!] \centering 
\includegraphics[width=0.7\textwidth]{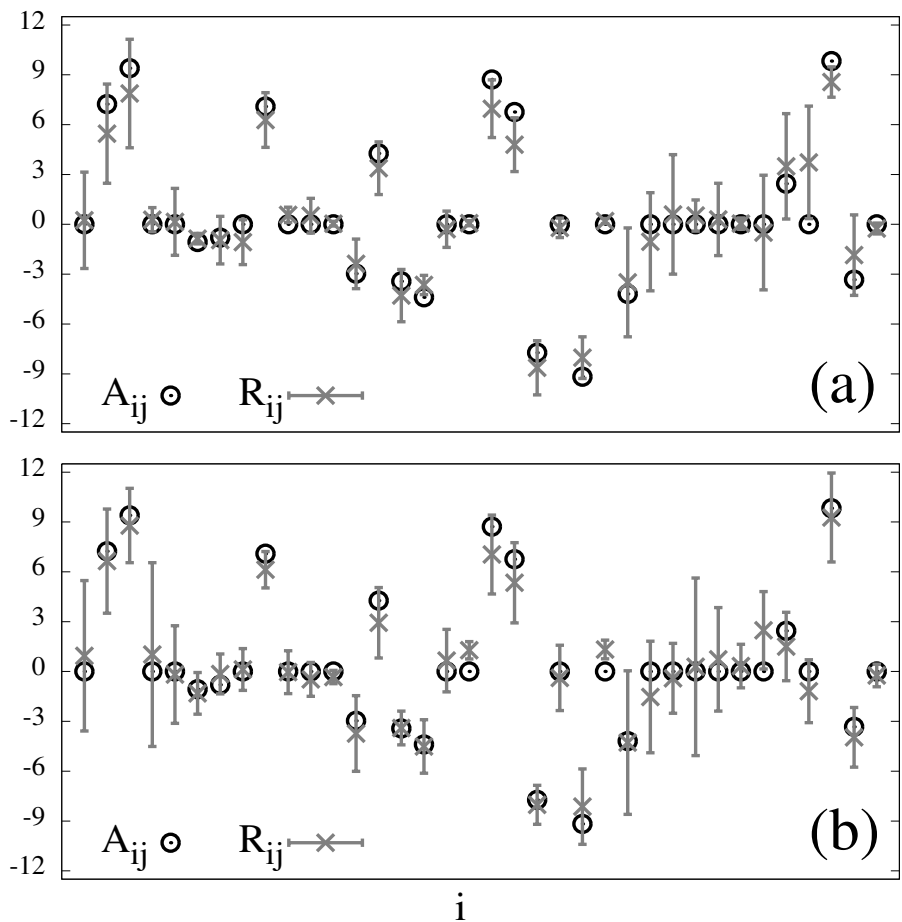}
\caption{Elements of the original $\A$ (circles), and the best reconstruction
(crosses), with the corresponding errorbars (first example). Model error and
observation error scenarios in (a) and (b), respectively.}  
\label{figure-6}   \end{figure}

In order to model the real applicability of our method, we test its robustness to possible violations of the 
initial assumptions, focusing on the first example (Fig.\ref{figure-1}a). 
We start with the scenario when the interaction
functions are not precisely known -- we assume a small mismatch in their
mathematical form (\textit{model error}). Instead of the original $\f = -x$ and
$\h = \tanh x$, we take $\f = -1.1x$ and $\h = \tanh(1.1x) + 0.1x$. The
measurements of $\Delta_T$ now cannot be expected to converge to zero.
Nevertheless, we apply the same procedure, and find (a weaker) correlation
between $\Delta_T$ and $\Delta_A$, as shown (by black dots) in Fig.\ref{figure-5}a. 
To see the worsening of the precision clearly, grey dots show the original 
non-perturbed scatter plot from Fig.\ref{figure-3}a. 
Dashed vertical line shows the part of the error $\Delta_T$ which is unavoidable due to 
the presence of the perturbation. We compute it as the difference between the original and
the perturbed interaction functions, averaged over the range of time series. We isolate this part of the 
trajectory error, since it is not due to the properties of our method. Its size indicates that the 
remaining part of the $\Delta_T$ is similar to the $\Delta_T$ occurring in the non-perturbed case. 
This demonstrates that our method works optimally even under perturbed conditions. The worsening of the reconstruction 
precision is what expected from the nature of the perturbation, meaning that our method makes 
no additional ``unexpected'' errors in the perturbed conditions.
The best reconstruction and the corresponding errorbars are computed as before and shown in Fig.\ref{figure-6}a. 
The errorbars are larger and the reconstruction precision worsens. Still, the essential fraction
of elements of $\A$ are within the respective errorbars. The decline of
precision is controllable, since it is clearly signalized by the size of the
errorbars. This could be used to generalize the method in the direction of
detecting the interaction functions as well. Each best $\R$ would be accompanied
by the best guesses for $f_i$ and $h_j$, meaning that different network
topologies, reproducing the data equally well, would come in combination with
different $f_i$ and $h_j$.

To test the second assumption of our theory, we take the time series to be not
precisely known due to \textit{observation errors}. Uncorrelated white noise of
intensity $\eta=0.4$ is added, perturbing each value of the time series. Instead
of the original data, we now consider one realization of the noisy data, as
illustrated in Fig.\ref{figure-2} (interaction functions are the original
ones). The central problem now is the computation of the derivatives, which are
extremely sensitive to the noise. We employ the Savitzky-Golay smoothing
filter~\cite{simonoff} as a standard technique of data de-noising, which allows
for a good derivative estimation. Since the time series are short, we apply the
smallest smoothing parameters. The reconstruction procedure is applied as
before, using smoothed derivatives to compute matrix $\mathbf{B}$ in
Eq.\ref{eq-2}. The scatter plot of $\Delta_T$ vs $\Delta_A$ is shown in Fig.\ref{figure-5}b, 
again compared with the original plot, and with the perturbation-induced unavoidable error indicated by the vertical line. The worsening of the precision 
is of a similar magnitude as in the model error scenario. The best reconstruction and the corresponding errorbars are reported in Fig.\ref{figure-6}b. 
Note that the precision is again correctly reflected by the size of the errorbars. In two cases from Fig.\ref{figure-6}, the
precision does not decline uniformly for all links. The analysis above shows
that our reconstruction method is reasonably robust to both model and
observation error. We found this robustness to be qualitatively independent of
the realization of both these errors.

\section*{Discussion and Conclusions}
We presented a method of reconstructing the topology of a general network from the dynamical time series with known interaction functions. Through
conceptually novel approach, our method is formulated as an inverse problem using linear systems formalism~\cite{joao}. Rather than relying on the
correlations between the observed variables, it is based on the correlations between the variables and their time derivatives. Our method involves two
important factors: it applies to the data that is relatively short, i.e. of the length comparable to the network size and to the characteristic time scale; and,
it yields the errorbars as a by-product, correctly reflecting the reconstruction precision.

On the other hand, our theory relies on knowing (at least approximately) both the dynamical model Eq.\ref{eq-1} and the interaction functions. While these assumptions might limit the immediate applicability of our method, our idea presents a conceptual novelty, potentially leading towards a more general and applicable reconstruction method. For example, we expect applicability in studies of interacting neurons in slices or cultures, where the properties of the individual neurons (i.e. functions $f$ and $h$) can be relatively well established, while the adjacency matrix is unknown. In contrast, the application to problems such as brain fMRI activity patterns, where even the existence of a dynamical model like Eq.\ref{eq-1} is questionable, appears at present not possible.

Our theory includes choosing the tunable observables $g$, which allow for the
reconstruction to be optimized within the constrains of any given data. The
question of constructing the optimal $g$ which extracts the \textit{maximal}
extractable information, remains open. Our algorithm can be reiterated: once the
1\% of the best $\R$-s are found, one can examine the functions $g$ leading to
those 1\%, and repeat the procedure, sampling only the neighboring portion of
the functional space. Alternatively, various evolutionary optimization
algorithms could be used~\cite{ja-srep}. An important factor for the method's
applicability is the dynamical regime behind the time series, which could be
regular (periodic) or chaotic (transiental). The former case is less
reconstructible, because of a poor coverage of the phase space. In particular, 
the synchronized dynamics, being essentially non-sensitive to the variations of the 
coupling coefficients, offers very little insight into the structure of the underlying 
network. Increasing the time series length is obviously of no help~\cite{preliminary}. 
In contrast, the latter case contains more network information, and is potentially more
reconstructible. Another issue is the applicability to large networks $N \ll 1$,
and in particular, the dependence of precision on relationship between $N$ and
$L$. This relates to the possibility of quantifying the network information
content of the available data. Relevant here is also the performance of our
method for varying types of network topologies (random, scalefree etc.). This is
a matter of ongoing research to be reported elsewhere.

Another limitation of our theory comes from the form of Eq.\ref{eq-1}. A similar theory could be
developed for alternative scenarios, such as $h$ specified by both source and
target nodes. The real challenge here are the networks with non-additive inter-node coupling (i.e., the dynamical contribution to the node $i$ is not a mere sum of neighbours' inputs). 
The key practical problem is that the mathematical forms of $f$
and $h$ are not (precisely) known for many real networks, although for certain
systems they can be inferred with a reasonable confidence~\cite{nelson,pigolotti}. Noise
always hinders the reconstruction, specially via derivative estimates. However,
longer time series not only bring more information, but also allow for a better
usage of smoothing. Finally, we note that the network reconstruction problem is
opposite of the network design problem. Our method could be employed to design a
network that displays given dynamics. However, while any network with $\Delta_T
\simeq 0$ solves the design problem, in the reconstruction theory this creates
the permanent issue of isolating the true network among those that exhibit
$\Delta_T \simeq 0$.

$\;\;$  \\\\
\noindent \textbf{Acknowledgments.} Work supported by the DFG via project FOR868, by ARRS via program P1-0383 ``Complex Networks'' and via project J1-5454 ``Unravelling Biological Networks''. Work also supported by Creative Core FISNM-3330-13-500033 ``Simulations'' funded by the European Union, The European Regional Development Fund. The operation is carried out within the framework of the Operational Programme for Strengthening Regional Development Potentials for the period 2007-2013, Development Priority 1: Competitiveness and research excellence, Priority Guideline 1.1: Improving the competitive skills and research excellence. We thank colleagues Michael Rosenblum and Mogens H. Jensen for useful discussions. \\\\

\noindent \textbf{Competing Financial Interests.} The authors declare no competing financial interests.\\\\

\noindent \textbf{Figure Legends:}
\begin{description}
\item[Figure 1] Adjacency matrices of two examined dynamical networks
\item[Figure 2] Example of timeseries produced by network Fig.~\ref{figure-1}a, including potential observation noise
\item[Figure 3] Scatter plots of errors $\Delta_T$ vs. $\Delta_A$ for the two studied cases
\item[Figure 4] Network reconstruction with errorbars for the two cases
\item[Figure 5] Scatter plots of errors $\Delta_T$ vs. $\Delta_A$ for the model error and observation error scenarios
\item[Figure 6] Network reconstruction with errorbars for the model error and observation error scenarios
\end{description}

\end{document}